%% file: iclr2026_conference.tex
\documentclass{article} 
\usepackage{iclr2026_conference,times}

\input{math_commands.tex}

\usepackage{hyperref}
\usepackage{url}
\usepackage{wrapfig}
\usepackage{graphicx} 
\usepackage{caption}
\usepackage{float}
\usepackage[most]{tcolorbox}
\usepackage{booktabs}
\title{Dial E for Ethical Enforcement: institutional VETO power as a governance primitive
}

 \iclrfinalcopy 

\author{
Subramanyam Sahoo\thanks{Correspondence: \texttt{\textbf{sahoo2vec@gmail.com}}}$^{1}$,  Vinija Jain$^{2,4}$,Aman Chadha$^{3,4}$, Divya Chaudhary$^{5}$\\[4pt]
$^{1}$Independent\\
$^{2}$Meta AI\\
$^{3}$AWS Generative AI Innovation Center, Amazon Web Services\\
$^{4}$Stanford University\\
$^{5}$Northeastern University, Seattle, WA, USA\\[6pt]
}

\begin{document}

\maketitle

\begin{abstract}
The persistent militarization of large reasoning models stems not from technical necessity but from governance arrangements that strip researchers of meaningful authority to refuse harmful transfers and deployments. Existing accountability mechanisms such as model cards and responsible AI statements operate as reputational signals detached from decision making architecture. We identify institutional veto power as a missing governance primitive: a formal authority to halt subsequent use or distribution of research when credible risks of weaponization emerge. Drawing on precedents in nuclear nonproliferation and biomedical ethics, the paper maps unprotected veto points across the research lifecycle, diagnose why compliance without enforceable constraints fails, and offer concrete institutional designs that embed veto authority while reducing the risk of political capture. The paper argues that communities most vulnerable to military uses must lead governance design, and that institutional veto power is a prerequisite for converting symbolic safeguards into enforceable responsibility and for achieving meaningful model disarmament.
\end{abstract}

\section{The Governance Paradox: Why Ethics Without Power Enables Militarization}

Contemporary AI governance suffers a paradox: widespread ethical awareness exists alongside near total incapacity to prevent harmful downstream uses. Current practices generate thorough risk knowledge, ethics statements, impact assessments, responsible AI codes, yet lack formal authority to block militarization or abusive deployment; this produces “organized irresponsibility,” where every actor can disclaim control while harms proceed. Unlike post WWII nuclear policy or biomedical oversight, which embedded refusal authority into institutions, AI governance treats loss of control after publication as inevitable even though militarization follows identifiable decision chains and discretionary institutional choices \cite{scharre_2023_four_battlegrounds}. We therefore recommend creating legally and procedurally protected institutional vetoes, targeted powers over transfer and deployment (not inquiry or publication), to condition use when credible risks such as militarization, surveillance misuse, or international law breaches are identified. Properly scoped, such refusal mechanisms would rebalance power toward researchers and affected communities and convert ethical knowledge into enforceable safeguards.

\begin{equation}
(K\land\neg A)\Rightarrow M\quad\text{and}\quad (K\land V\land R)\Rightarrow\neg M\quad\Rightarrow\quad P(M\mid V)\ll P(M\mid\neg V)
\label{eq:compressed_paradox}
\end{equation}

\noindent\textbf{Notation.}
Here, $K$ denotes ethical knowledge produced by governance practices, $A$ denotes institutional authority to refuse or block use, $M$ denotes militarization or abusive deployment, $R$ denotes the identification of a credible risk, and $V$ denotes a legally and procedurally protected veto over transfer or deployment.

\section{Related Work: Veto Power in Governance}

Current AI governance literature emphasizes either technical safety mechanisms (alignment, interpretability) or procedural ethics (review boards, impact assessments) \cite{sahoo2025votemultistakeholderframeworklanguage}. Yet this work treats militarization as a downstream consequence amenable to ethics mitigation, not as an architectural problem requiring authority redistribution. The governance literature in other domains—nuclear nonproliferation, bioethics, labor rights—demonstrates that constraining specific uses requires veto authority, not merely visibility \cite{Crawford+2021,kalluri2023surveillanceaipipeline}. Our contribution bridges these literatures: \textit{we argue that AI militarization is fundamentally a governance problem, not an ethics problem, and therefore requires mechanisms (veto power) that other high risk domains have already institutionalized.} Existing proposals for ``Responsible AI'' governance lack enforcement mechanisms precisely because they assume voluntary institutional compliance; veto power reverses this assumption, making restraint collectively enforceable. The dual use literature identifies risks but not solutions. We propose an operational mechanism that could actually arrest militarization trajectories at decision points where prevention remains possible. \textbf{By framing veto power as a governance primitive—a basic building block that other domains have proven workable—we shift discourse from ``how can we encourage restraint'' (eternally failing) to ``how can we make restraint structurally rational'' (institutionally achievable).}

\section{Mapping Discretionary Militarization: Unprotected Veto Points}

\begin{figure}[H]
    \centering
    \includegraphics[width=0.95\linewidth]{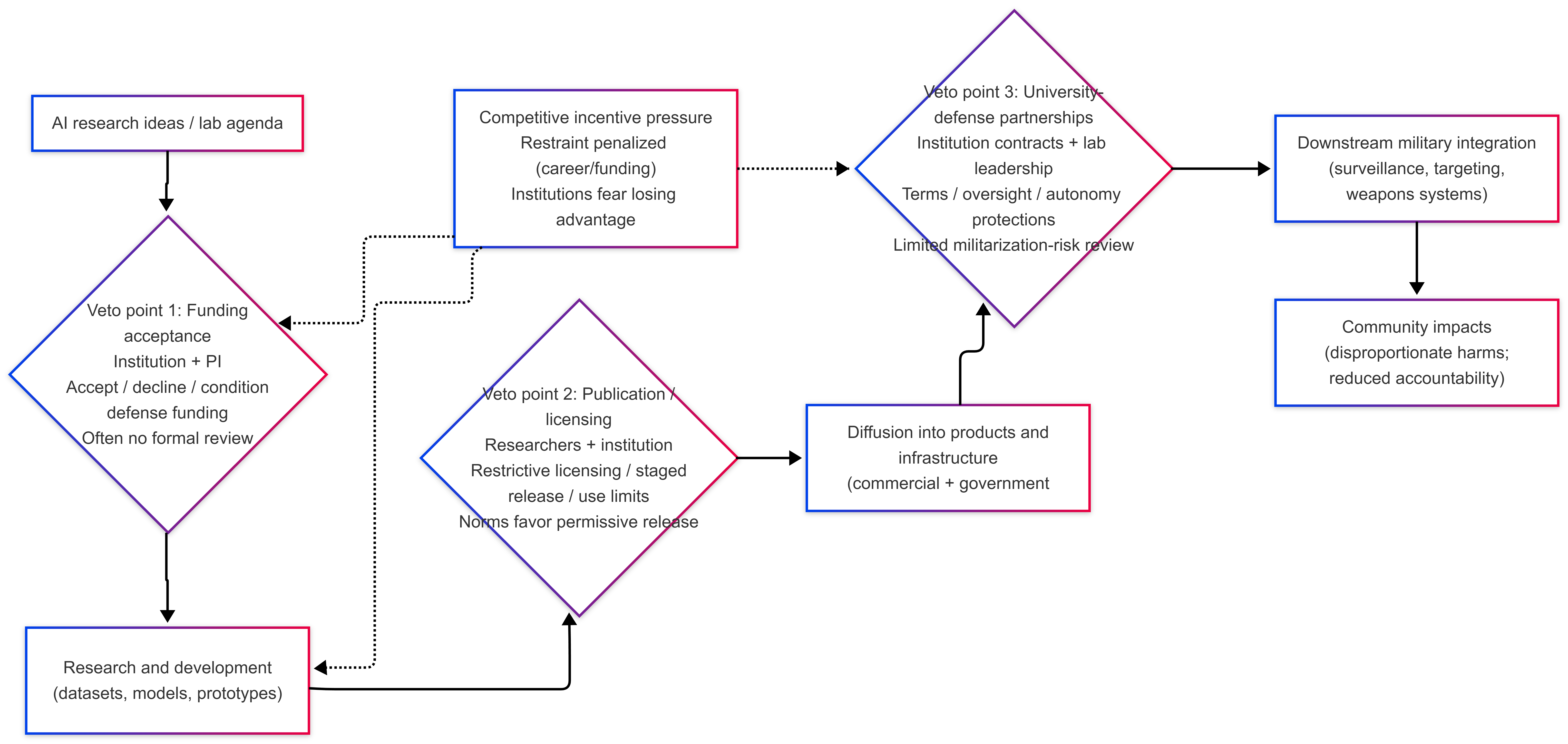}
    \caption{Workflow of discretionary militarization in AI research, highlighting institutional veto points and the competitive incentive pressures that reinforce downstream military integration and community harms.}
    \label{fig:placeholder}
\end{figure}

Militarization of AI is driven by routine institutional choices that act as veto points: acceptance of military research funding, decisions about publication and licensing, and formal partnerships between universities and defense firms \cite{lamparth2024humanvsmachinebehavioral}. Each of these moments could be governed differently. Institutions could require review before accepting military grants, adopt restrictive licenses or license conditions that bar surveillance use, delay publication pending risk assessment, or subject partnership proposals to independent 3rd party oversight. Instead these choices are treated as individual researcher preferences, so universities routinely accept funding and sign agreements without deliberation about downstream military integration and researchers who resist face career penalties and loss of resources\cite{khlaaf2024mindgapfoundationmodels}.

These veto points share a structural defect. Authority is exercised without meaningful participation from affected communities or institutional accountability, producing a collective action failure in which restraint is socially preferable but individually costly. Actors who might exercise restraint are undercut by competitors that do not. Addressing this requires formal governance mechanisms that distribute authority, mandate risk review, and protect those who refuse military integration so that the collectively rational choice of restraint becomes individually viable \cite{fli_autonomous_weapons_2016_online}, 2016.

\section{From Ethics Without Veto to Enforceable Governance}
Current ethics frameworks fail for structural reasons that better guidelines cannot fix. Responsibility is dispersed across researchers, universities, philanthropists, tech transfer offices, and procurement chains, so no actor holds both the authority and obligation to stop militarization—and each can deflect blame when harms occur. Voluntary restraint also produces adverse selection: institutions that decline military funding, impose use restrictions, or protect research autonomy pay real costs while less restrained competitors gain advantage, pushing trajectories toward militarization regardless of ethical exhortation . Finally, ethics processes \textit{enable laundering}: documentation of “due consideration” can psychologically and institutionally substitute for changed decisions, allowing actors to treat deliberation as discharge of responsibility even when outcomes remain unchanged.

These failure modes require architectural change. We propose institutional veto power: governance that makes ethical judgments binding, locates decision authority with those exposed to consequences, and turns restraint from an individually punished choice into a collectively enforceable rule. Veto is not arbitrary refusal but narrowly triggered by predefined thresholds—e.g., credible military integration, surveillance targeting marginalized groups, or violations of international law—paired with collective decision-making, appeal procedures, and anti-retaliation protections. Crucially, this does not require new treaties: universities can revise contracts and create veto committees; funding parties can condition grants on enforceable safeguards; conferences and repositories can require disclosure and enforce use conditions; and professional communities can adopt binding, public norms that incrementally embed veto governance within existing institutions \cite{doi:10.1177/17470161241267782}.

\section{Centering Justice: Veto Power Requires Marginalized Community Leadership}

\textbf{A core governance question concerns who determines which militarization risks justify veto authority and whose interests such decisions ultimately serve.} Contemporary debates remain concentrated within elite academic and policy institutions, largely dominated by socially privileged actors, while communities that experience the most severe downstream harms remain structurally excluded. These include populations subjected to automated targeting and large scale civilian harm, minorities exposed to pervasive surveillance infrastructures, and migrant groups whose autonomy is constrained by biometric identification systems. This exclusion is not only normatively problematic but also epistemically limiting. Affected communities possess situated knowledge about system deployment, harm salience, and justice relevant tradeoffs that cannot be adequately represented through advisory consultation within externally designed governance frameworks \cite{bengio2025internationalaisafetyreport}. From a policy perspective, justice therefore requires operational community leadership rather than symbolic inclusion. This entails binding decision authority. In practice, veto bodies governing surveillance systems should include civil rights advocates and representatives from surveilled communities with decision making power equal to that of technical experts and ethicists. Similarly, governance arrangements for autonomous weapons should be co designed with researchers and affected populations in regions of deployment, rather than developed within wealthy institutions and later exported as policy. Research funding decisions shaping militarized contexts should formally incorporate community voices from those regions, with mechanisms that allow community objections to block projects rather than merely inform them. This implies structural reform toward community controlled governance in which affected populations directly hold veto authority.

Militarization and marginalization are deeply interlinked. Military violence disproportionately targets already marginalized groups, AI systems scale and automate existing targeting logics, and data extracted from these communities is used to train systems later deployed against them. Within this context, veto authority should function bidirectionally. Beyond constraining militarized trajectories, governance mechanisms should actively enable and protect research oriented toward peace and justice. Such efforts should be institutionally facilitated rather than incidentally tolerated. Veto based governance therefore performs a dual role: it blocks pathways toward weaponization while simultaneously safeguarding and accelerating peace oriented innovation. When affected communities hold veto authority, they can refuse militarized research agendas while directing institutional resources toward technologies that advance collective safety, autonomy, and long term justice\cite{khlaaf2024mindgapfoundationmodels,9795193}.

\begin{figure}[H]
    \centering
    \includegraphics[width=1\linewidth]{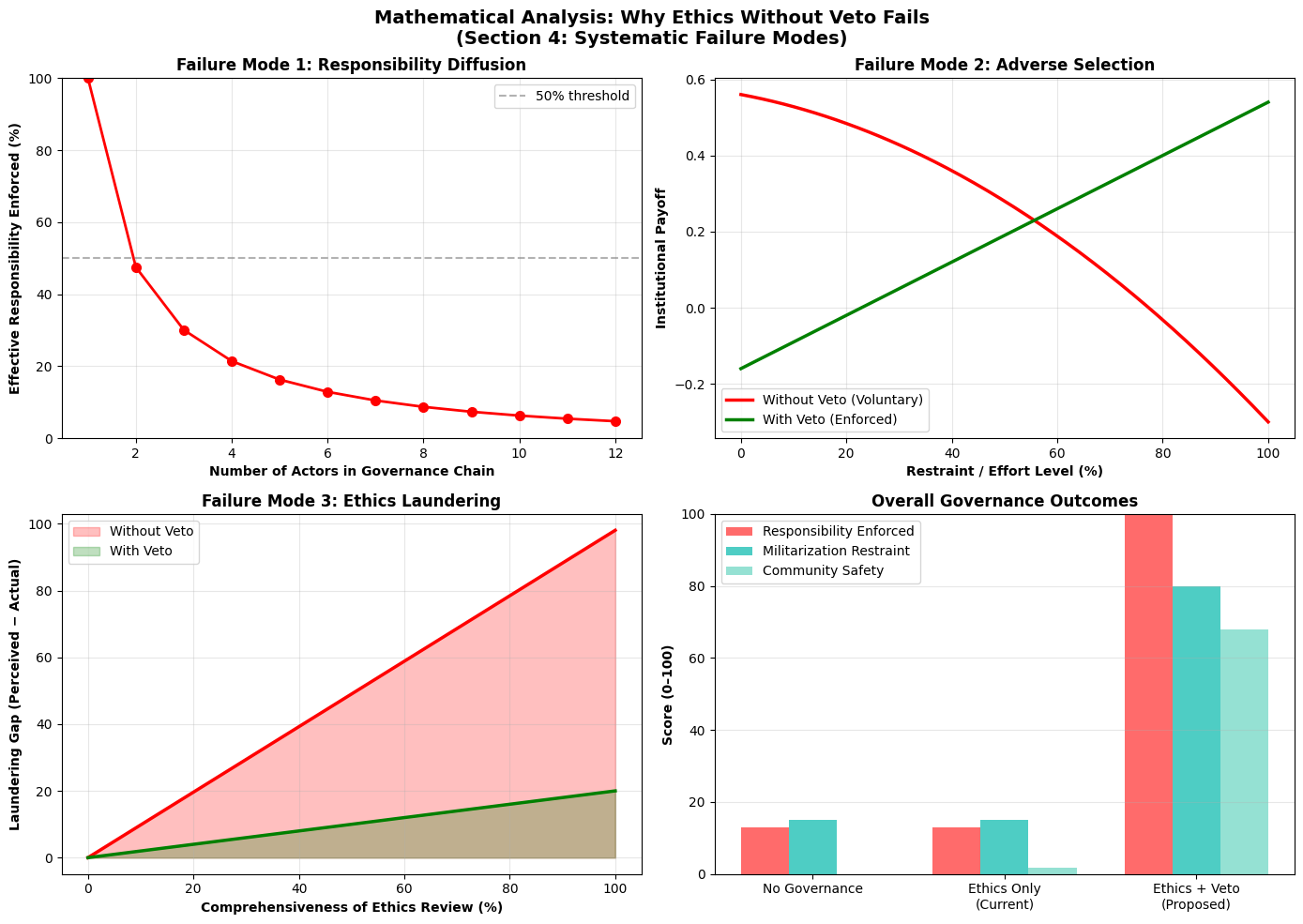}
    \caption{Why ethics without veto power fails to constrain militarization. (a) Responsibility diffusion: enforceable responsibility falls as governance chains lengthen. (b) Adverse selection: competition drives low restraint without enforcement; veto stabilizes higher restraint. (c) Ethics laundering: documentation inflates perceived responsibility without intervention absent enforceable authority. (d) Derived outcomes (responsibility enforced, restraint, community safety) under no governance, ethics-only, and ethics+veto regimes.}
    \label{fig:placeholder}
\end{figure}

\section{Limitations and Future Work}
Veto power can make decisions about militarized research enforceable, but it does not by itself fix deeper power imbalances. It will not redistribute research funding to marginalized communities, democratize university governance, dismantle the defense industry that shapes priorities, or automatically hold researchers accountable, and implementation will encounter institutional inertia and the risk of capture by hostile actors. Empirical work is needed: document and analyze real veto decisions to build precedent, run longitudinal studies of research trajectories and weaponization timelines, center leadership and partnership with communities in the Global South who are directly affected, and study how veto mechanisms fit with export controls, international treaties, supplier transparency, and legal accountability so veto power does not become another way to ease guilt while leaving power structures intact.

\section{Conclusion: The Choice Before Research Communities}

We face a choice between continuing with governance that allows ethics to flourish as performance while harm accumulates and responsibility diffuses, or moving toward collective, binding restraint where decisions about research militarization are accountable and constrained. Veto power offers a structural path for that shift, so researchers need not rely on lone moral courage, institutions need not suffer competitive disadvantage for practicing restraint, and affected communities gain voice in decisions about their lives. Performative ethics are over; enforceable, community-led responsibility to prevent militarization and promote peace must begin.

\section*{LLM Usage Disclosure}

This work employed large language models in a supporting capacity. Specifically, we used Claude 4.5 Haiku (Anthropic, 2024) for the following roles:

\paragraph{Writing Assistance.} The LLM was just asked to remove artificial Overleaf latex errors.

\paragraph{Limitations of LLM Use.} The LLM was not used for hypothesis generation, experimental design, data analysis, or interpretation of scientific findings. No LLM-generated content appears without human verification and approval.

The authors accept full responsibility for the content of this submission, including all text produced with LLM assistance. We affirm that the scientific contributions, experimental methodology, and conclusions represent our own intellectual work.

\bibliography{iclr2026_conference}
\bibliographystyle{iclr2026_conference}

\appendix
\section*{Appendix}

\section{Toy model: why ethics without veto power fails}

\subsection{Notation}
Let $n$ be the number of actors in a governance chain.
Let restraint be $r\in[0,1]$ and documentation (ethics review) be $D\in[0,1]$.
Let baseline harm be $H_0>0$.
Let $\alpha\in(0,1)$ denote a per-actor coordination probability.
We use a normalized total responsibility $R_{\mathrm{tot}}$.

\subsection{Failure mode 1: responsibility diffusion}
\begin{align}
R_{\mathrm{ind}}(n) &= \frac{R_{\mathrm{tot}}}{n}, \label{eq:Rind}\\
p(n) &= \alpha^{\,n-1}, \label{eq:pn}\\
\rho(n) &= \frac{R_{\mathrm{ind}}(n)\,p(n)}{R_{\mathrm{tot}}}
       = \frac{\alpha^{\,n-1}}{n}. \label{eq:rho}
\end{align}
Here $\rho(n)\in[0,1]$ is the effective enforceability fraction.
With a binding veto point, enforceability is consolidated:
\begin{align}
\rho_{\mathrm{veto}} &= 1. \label{eq:rhoveto}
\end{align}

\subsection{Failure mode 2: adverse selection under voluntary restraint}
Let $\bar r$ be the average restraint of competitors. A stylized payoff without veto is
\begin{align}
U_{\mathrm{noveto}}(r;\bar r)
&= \beta(1-r)\bar r + \gamma r - \kappa r^2, \label{eq:Unoveto}
\end{align}
where $\beta>0$ captures militarization/funding advantage, $\gamma\ge 0$ captures ethical signaling,
and $\kappa>0$ is the cost of restraint.

A best response and symmetric equilibrium are defined by
\begin{align}
\mathrm{BR}(\bar r) &= \arg\max_{r\in[0,1]} U_{\mathrm{noveto}}(r;\bar r), \label{eq:BR}\\
r^\ast_{\mathrm{noveto}} &= \mathrm{BR}(r^\ast_{\mathrm{noveto}}). \label{eq:req}
\end{align}

With veto enforcement at level $\bar R\in[0,1]$,
\begin{align}
r^\ast_{\mathrm{veto}} &= \bar R. \label{eq:rveto}
\end{align}

\subsection{Failure mode 3: ethics laundering and intervention}
Perceived responsibility increases with documentation:
\begin{align}
P(D) &= 100D. \label{eq:perceived}
\end{align}
Actual harm prevented depends on intervention $I\in[0,1]$:
\begin{align}
A(I) &= H_0 I. \label{eq:actual}
\end{align}
The ethics-laundering gap is
\begin{align}
G(D,I) &= P(D) - A(I). \label{eq:gap}
\end{align}

We link documentation to intervention by enforceability and equilibrium restraint:
\begin{align}
I &= D \cdot \rho \cdot r^\ast, \label{eq:intervention}
\end{align}
where $\rho\in\{\rho(n),\rho_{\mathrm{veto}}\}$ and $r^\ast\in\{r^\ast_{\mathrm{noveto}},r^\ast_{\mathrm{veto}}\}$.

Resulting harm is
\begin{align}
H &= H_0(1-I). \label{eq:harm}
\end{align}

\subsection{Derived summary metrics}
We report:
\begin{align}
\text{ResponsibilityEnforced} &= 100\rho, \label{eq:m1}\\
\text{MilitarizationRestraint} &= 100r^\ast, \label{eq:m2}\\
\text{AdverseSelectionScore} &= 100(1-r^\ast), \label{eq:m3}\\
\text{HarmToCommunities} &= H, \qquad
\text{CommunitySafety}=100-H. \label{eq:m4}
\end{align}

\begin{tcolorbox}[
  title={Model parameters (“magic numbers”) and scenario defaults},
  colback=yellow!6,
  colframe=orange!70!black,
  fonttitle=\bfseries,
  boxrule=0.9pt,
  arc=2mm,
  left=2mm,
  right=2mm,
  top=1mm,
  bottom=1mm
]
\small 
\setlength{\tabcolsep}{4pt} 

\noindent\textbf{Python parameter dataclasses used in the replication code.}

\vspace{0.6em}
\noindent\textbf{ModelParams (default coefficients)}

\medskip
\begin{minipage}{\linewidth}
\begin{tabular}{@{} p{0.20\linewidth} p{0.12\linewidth} p{0.66\linewidth} @{}}
\toprule
Symbol / name & Default & Meaning \\
\midrule
$\alpha$ (\texttt{alpha}) & $0.95$ & Per-actor coordination probability (diffusion model) \\
$R_{\mathrm{tot}}$ (\texttt{total\_responsibility}) & $100$ & Responsibility normalization \\
$\beta$ (\texttt{beta}) & $0.8$ & Militarization / funding advantage coefficient \\
$\gamma$ (\texttt{gamma}) & $0.3$ & Ethical prestige / signaling coefficient \\
$\kappa$ (\texttt{kappa}) & $0.6$ & Quadratic cost of restraint \\
$q$ (\texttt{quality\_coeff}) & $0.7$ & Prestige from research effort/quality (with veto) \\
$c_v$ (\texttt{veto\_cost\_coeff}) & $0.2$ & Shared cost of enforced governance (with veto) \\
$H_0$ (\texttt{baseline\_harm}) & $100$ & Baseline harm scale if no effective intervention \\
\bottomrule
\end{tabular}
\end{minipage}

\vspace{0.8em}
\noindent\textbf{GovernanceScenario (default scenario fields)}

\medskip
\begin{minipage}{\linewidth}
\begin{tabular}{@{} p{0.20\linewidth} p{0.12\linewidth} p{0.66\linewidth} @{}}
\toprule
Field & Default & Meaning \\
\midrule
\texttt{name} & --- & Scenario label \\
$n$ (\texttt{num\_actors}) & $6$ & Number of actors in governance chain \\
$D$ (\texttt{documentation\_level}) & $0.0$ & Ethics documentation/review level, $D\in[0,1]$ \\
\texttt{has\_veto} & \texttt{False} & Whether binding veto enforcement exists \\
$\bar R$ (\texttt{veto\_enforcement\_level}) & $0.0$ & Enforced restraint level if veto exists, $\bar R\in[0,1]$ \\
\bottomrule
\end{tabular}
\end{minipage}
\end{tcolorbox}

\section{VETO MECHANISMS IN COMPARATIVE GOVERNANCE CONTEXTS}

This appendix establishes that veto power is not a novel or untested governance approach. Rather, it represents a well-established mechanism that has successfully constrained dangerous technologies in other high-risk domains. We examine three historical and contemporary examples: nuclear nonproliferation regimes, institutional review boards in biomedical research, and corporate compliance in export control regimes.

\section*{B.1 Nuclear Nonproliferation Regimes}

\subsection*{Historical Context and Necessity}

The Nuclear Non-Proliferation Treaty (NPT), negotiated in 1968 and entered into force in 1970, emerged from recognition that uncontrolled nuclear weapons proliferation posed existential risk. Following the development of nuclear weapons by the Soviet Union (1949) and subsequent weapons development by additional states, the international community recognized that preventing further proliferation required not merely ethical exhortation but institutional mechanisms with enforcement authority.

The NPT created a legal framework establishing three categories of states: Nuclear Weapons States (the five permanent UN Security Council members), Non-Nuclear Weapons States, and threshold states capable of developing weapons. The treaty created asymmetric obligations: Nuclear Weapons States committed (in principle) to disarmament; Non-Nuclear Weapons States committed to forgo weapons development; and all states committed to allowing international inspection.

\subsection*{The International Atomic Energy Agency (IAEA): Veto Power Through Verification}

The IAEA, established in 1957 and strengthened by the NPT framework, represents the institutional embodiment of veto power in nuclear governance. The IAEA possesses several critical authorities that function analogously to the veto mechanisms proposed in this paper:

\begin{enumerate}
\item \textbf{Inspection Authority}: The IAEA maintains the right to inspect nuclear facilities in Non-Nuclear Weapons States at declared locations and, in many cases, at undeclared locations suspected of weapons development. These inspections are not advisory; they are mandatory conditions of NPT membership. The IAEA inspectorate can refuse to declare a state ``in compliance'' if inspections reveal violations or lack of transparency. This refusal to certify compliance effectively blocks a state's standing in the international community and triggers scrutiny and sanctions.

\item \textbf{Materials Tracking}: The IAEA tracks nuclear materials (uranium, plutonium, enriched fuel) as they move through the fuel cycle. Unexplained disappearance of fissile materials or transfer to undeclared facilities triggers investigation and can lead to referral to the UN Security Council for sanctions.

\item \textbf{Technology Transfer Restrictions}: The IAEA maintains safeguards on sensitive nuclear technology, restricting the circumstances under which uranium enrichment technology, plutonium reprocessing technology, or other weapons-relevant capabilities may be transferred internationally. These restrictions function as gatekeeping mechanisms preventing proliferation.

\item \textbf{Enforcement Through Referral}: While the IAEA itself lacks enforcement authority (it cannot impose sanctions), persistent violations trigger mandatory referral to the UN Security Council, which can impose sanctions, travel bans, asset freezes, and military intervention. This creates automatic escalation consequences for violation.
\end{enumerate}

\subsection*{Why Nuclear Nonproliferation Succeeded (Partially)}

The nuclear nonproliferation regime has not eliminated weapons proliferation---the number of nuclear weapons states has increased from 5 to 9 since the NPT entered into force. However, it has constrained proliferation significantly below models predicting 20-30+ weapons states by the 1980s. The regime succeeded to this degree because:

\begin{itemize}
\item \textbf{Verification was mandatory, not voluntary}: States could not choose whether to allow inspections; inspections were binding conditions of participation.

\item \textbf{Veto points were clearly defined}: Uranium enrichment, plutonium reprocessing, and weapons-relevant technology transfer were identified as specific decision points where control could be exercised.

\item \textbf{Consequences were automatic}: Violations triggered referral to the Security Council, which imposed sanctions without requiring additional political negotiation.

\item \textbf{Enforcement was international}: No single state could unilaterally shield violators; enforcement required coordination across multiple states, reducing the likelihood of capture by individual interests.
\end{itemize}

\subsection*{Limitations and Implications for AI Governance}

The nuclear nonproliferation regime also reveals limitations relevant to AI governance design:

\begin{itemize}
\item \textbf{Enforcement gaps}: States with nuclear weapons or Security Council vetoes can shield themselves or allies from referral and sanctions, limiting enforcement against major powers. This suggests that AI veto governance should include mechanisms preventing powerful actors from shielding themselves.

\item \textbf{Inspection difficulty}: Detecting undeclared nuclear programs remains challenging; the IAEA's detection of\textbf{ Iraq's covert program (1991) and Iran's nuclear work} depended partly on intelligence agencies and whistleblowing rather than purely technical verification.

\item \textbf{Treaty withdrawal}: States can withdraw from the NPT with limited consequences; North Korea withdrew in 2003. This suggests that voluntary governance frameworks have inherent fragility.
\end{itemize}

\subsubsubsection{IAEA Authority Mechanisms Relevant to AI}

The IAEA's specific operational mechanisms offer concrete models for AI veto governance (see Table~\ref{table:iaea_analogs}).

\begin{table}[h]
\centering
\small
\setlength{\tabcolsep}{4pt}
\begin{tabular}{p{0.27\columnwidth} p{0.30\columnwidth} p{0.35\columnwidth}}
\hline
\textbf{Nuclear Mechanism} & \textbf{Purpose} & \textbf{AI Analog} \\
\hline
Mandatory inspections &
Verify compliance with restrictions &
Repository audits verify permitted use restrictions. \\

Materials accounting &
Track fissile materials through the fuel cycle &
Artifact-tracking logs record access and usage. \\

Technology controls &
Restrict transfer of enrichment technology &
License restrictions limit downstream licensing. \\

Referral to higher authority &
Escalate violations to political bodies &
Funding-agency enforcement ties compliance to grants. \\
\hline
\end{tabular}
\caption{IAEA mechanisms and AI governance analogs}
\label{table:iaea_analogs}
\end{table}

\section*{B.2 Institutional Review Boards in Biomedical Research}

\subsection*{Historical Context: The Tuskegee Case and Regulatory Response}

Institutional Review Boards (IRBs) emerged from historical horror. The Tuskegee Syphilis Study (1932--1972), conducted by the U.S. Public Health Service, enrolled African American men with syphilis into a study ostensibly to track the natural history of the disease. In reality, participants were not informed of their diagnosis, were not offered treatment even after penicillin became available (1943), and were subjected to deceptive medical procedures. The study continued for 40 years, resulting in preventable illness and death among study subjects. The scandal, publicly exposed in 1972, triggered comprehensive regulatory reform.

The regulatory response, codified in the Belmont Report (1979) and subsequent federal regulations (45 CFR 46), established that institutions receiving federal research funding must establish Institutional Review Boards with authority to review, approve, and halt research involving human subjects. IRBs operate at the threshold of research: before researchers begin studies, proposed research must be reviewed by a committee including scientists, ethicists, community representatives, and legal experts.

\subsection*{IRB Authority and Scope}

IRBs possess genuine veto authority within their defined scope:

\begin{enumerate}
\item \textbf{Pre-Implementation Review}: IRBs review research proposals before implementation. This timing is critical: it allows IRBs to block research before resources are committed and before recruitment proceeds.

\item \textbf{Binding Authority}: IRB decisions are binding. Researchers cannot proceed without IRB approval; attempting to do so violates federal regulation and triggers sanctions against the researcher and institution.

\item \textbf{Protected Refusal}: Federal regulation protects IRBs and researchers from retaliation for IRB decisions. Institutions cannot punish researchers whose protocols are rejected.

\item \textbf{Continuing Oversight}: IRBs retain authority throughout a study, conducting ongoing review of protocols, reviewing adverse events, and requiring corrective action if risks exceed expected levels.
\end{enumerate}

\subsection*{Why IRBs Succeeded in Constraining Research Harms}

IRBs have demonstrably constrained research harms compared to the pre-IRB era. Studies now rarely involve the level of deception and harm that characterized Tuskegee or Nazi medical experiments. This success stems from:

\begin{itemize}
\item \textbf{Authority at the decision point}: By reviewing before research begins, IRBs operate at the moment when research trajectories can be most easily altered.

\item \textbf{Protected collective deliberation}: IRBs bring multiple perspectives (scientific, ethical, community) to bear on proposals.

\item \textbf{Routine, normalized process}: IRB review is standard practice in biomedical research, institutionalized rather than extraordinary.

\item \textbf{Regulatory backing}: IRBs are backed by federal regulation and funding consequences. Institutions that fail to maintain functional IRBs lose federal research funding eligibility.
\end{itemize}

\subsection*{Limitations of IRBs}

IRBs also reveal significant limitations relevant to AI governance design:

\begin{enumerate}
\item \textbf{Scope Limitation}: IRBs focus exclusively on risks to research subjects---direct, identifiable individuals participating in studies. They have no authority over risks to populations not in the study (third-party harms), societal harms, or dual-use concerns.

\item \textbf{Implementation Versus Use}: IRBs review the research process itself but do not review downstream uses of research findings. Once research is published, IRBs have no authority to restrict how others use the knowledge.

\item \textbf{Variable Quality and Authority}: IRB quality varies significantly. Underfunded IRBs with limited expertise may fail to identify harms or may be captured by institutional interests.

\item \textbf{Institutional Capture}: While formal protections exist against retaliation, institutional culture can undermine IRB authority.
\end{enumerate}

\subsubsubsection{IRB Mechanisms Applicable to AI Veto Governance}

See Table~\ref{table:irb_analogs} for specific operational mechanisms applicable to AI governance.

\begin{table}[h]
\centering
\small
\setlength{\tabcolsep}{4pt}
\begin{tabular}{p{0.27\columnwidth} p{0.30\columnwidth} p{0.40\columnwidth}}
\hline
\textbf{IRB Mechanism} & \textbf{Purpose} & \textbf{AI Analog} \\
\hline
Pre-implementation review &
Assess protocols before initiation &
Pre-licensing review prior to publication. \\

Institutional requirement &
Mandate review as condition of funding &
Funding conditional on veto governance. \\

Protected refusal &
Protect from retaliation &
Anti-retaliation safeguards for objectors. \\

Continuing oversight &
Review throughout implementation &
Ongoing deployment monitoring. \\

Documented procedures &
Transparent criteria and procedures &
Predefined veto triggers and workflows. \\

Pluralistic composition &
Multiple perspectives in review &
Diverse veto bodies with varied expertise. \\
\hline
\end{tabular}
\caption{IRB mechanisms and AI governance analogs}
\label{table:irb_analogs}
\end{table}

\section*{B.3 Corporate Compliance and Export Control Regimes}

\subsection*{Export Control Framework and Dual-Use Technology Governance}

Export control regimes represent a third governance domain where veto-like mechanisms have been operationalized at scale. The International Traffic in Arms Regulations (ITAR), the Export Administration Regulations (EAR), and parallel regimes in other countries, establish that certain technologies cannot be transferred internationally without government authorization. These regimes recognize that technology transfer represents a decision point where control can and should be exercised.

\subsection*{Specific Export Control Mechanisms}

\begin{enumerate}
\item \textbf{Commodity Identification}: Export control regimes maintain detailed lists specifying which technologies are controlled. Companies must determine whether their products fall within controlled categories and, if so, comply with licensing requirements.

\item \textbf{Licensing Requirements}: Transfer of controlled items requires government authorization. Companies submit license applications specifying the technology, the proposed recipient, the intended end use, and the end user.

\item \textbf{End-Use Commitments}: Licensees must commit to specific end uses and are prohibited from transferring items to third parties without subsequent authorization.

\item \textbf{Verification and Audit}: Companies maintaining export licenses are subject to government audits verifying compliance with license conditions.

\item \textbf{Penalties for Violation}: Unauthorized export of controlled items triggers significant penalties: civil penalties (fines up to \$300,000 per violation), criminal penalties (up to 20 years imprisonment for willful violation).
\end{enumerate}

\subsection*{Why Export Controls Succeeded}

Export control regimes have demonstrably constrained the transfer of sensitive military technologies to hostile regimes, though imperfectly. Successes include:

\begin{itemize}
\item \textbf{Slowed weapons development}: Export controls on advanced semiconductor manufacturing equipment delayed China's development of advanced semiconductors, constraining military applications.

\item \textbf{Prevented technology cascades}: Export controls on encryption technology slowed proliferation to adversarial states.

\item \textbf{Deterred proliferation}: Companies face significant penalties for unauthorized export; this creates incentives to comply with restrictions.
\end{itemize}

\subsubsubsection{Export Control Mechanisms Applicable to AI Veto Governance}

See Table~\ref{table:export_analogs} for specific export control mechanisms applicable to AI governance.

\begin{table}[h]
\centering
\small
\setlength{\tabcolsep}{3pt}
\renewcommand{\arraystretch}{1.08}
\begin{tabular}{@{} p{0.22\columnwidth} p{0.28\columnwidth} p{0.45\columnwidth} @{}}
\hline
\shortstack{\textbf{Export Control}\\\textbf{Mechanism}} &
\textbf{Purpose} &
\textbf{AI Analog} \\
\hline
Commodity lists &
Specify controlled tech\-nologies &
Define dual-use research categories and flagged capa\-bilities. \\

License requirements &
Authorize transfers &
Pre-licensing review prior to transfers. \\

End-use commitments &
Restrict recipient uses &
Use restrictions encoded in licenses and agree\-ments. \\

Verification and audit &
Monitor compliance &
Deployment attesta\-tions, audits, and provenance checks. \\

Penalties for violation &
Impose consequences &
Funding ineligibility, sanctions, and liability. \\
\hline
\end{tabular}
\caption{Export control mechanisms and AI governance analogs}
\label{table:export_analogs}
\end{table}

\subsection*{Integration Across Domains: Lessons for AI Veto Governance}

Examining nuclear nonproliferation, IRBs, and export controls together reveals common principles applicable to AI governance:

\begin{enumerate}
\item \textbf{Veto operates at decision points, not at knowledge generation}: Nonproliferation restricts uranium enrichment, not uranium chemistry. IRBs restrict research implementation, not inquiry. Export controls restrict transfer, not publication. In each case, veto operates at a specific decision point where control remains possible without censoring knowledge itself.

\item \textbf{Authority must be binding and enforceable}: Successful governance regimes embed veto authority in formal decision-making structures with binding effect. Recommendations, guidelines, and voluntary frameworks consistently fail to constrain motivated actors.

\item \textbf{Consequences must be material and automatic}: Violations trigger tangible consequences (sanctions, loss of funding, criminal liability) rather than relying on reputation or goodwill. Automatic consequences prove more effective than discretionary penalties.

\item \textbf{Governance is politically difficult but necessary}: All three regimes faced and continue to face resistance from actors who benefit from unconstrained use of technology. Yet all three have achieved partial success at constraining dangerous applications.
\end{enumerate}


\section{IMPLEMENTATION PATHWAYS - DETAILED MECHANISMS}

Veto power becomes operationally effective when embedded in concrete institutional arrangements. This section details five specific implementation pathways through which veto governance can be implemented using existing institutional structures. Critically, these pathways do not require new governmental authority, new treaties, or new institutions.

\section*{C.1 University Contract Modifications: Embedding Veto in Technology Transfer}

\subsection*{Current Landscape}

Universities currently manage significant research portfolios, including dual-use research with foreseeable military applications. Technology transfer offices (TTOs) license research to downstream organizations, negotiate research partnerships, and manage intellectual property. These decisions are typically made without institutional deliberation about downstream military integration.

\subsection*{Mechanism: Veto Clauses in Licensing Agreements}

Universities can embed veto authority by modifying standard licensing agreements to include contractual clauses specifying:

\begin{enumerate}
\item \textbf{Use Restrictions}: Licenses can specify that research will not be transferred for military use, surveillance targeting of marginalized groups, or deployment in violation of international humanitarian law.

\item \textbf{Institutional Review Requirement}: Licenses can require that transfers to third parties undergo institutional review before the transfer can occur.

\item \textbf{Deployment Reporting}: Licenses can require that licensees report on deployment contexts, provide documentation of end uses, and notify the university if deployment deviates from licensed purposes.

\item \textbf{Indemnification for Violation}: Licenses can specify that licensees assume legal liability if they deploy research in violation of license restrictions.

\item \textbf{Right to Audit}: Licenses can grant universities the right to conduct audits of licensees' facilities and deployment contexts, verifying compliance with use restrictions.
\end{enumerate}

\subsection*{Implementation Example}

Suppose a university develops facial recognition technology with clear surveillance applications. Under veto governance, a modified licensing agreement might specify:

\begin{quote}
``\textit{The research may not be deployed for mass surveillance targeting political opponents, ethnic minorities, or other vulnerable populations without university approval. Licensee must report annually on deployment contexts. University maintains the right to audit licensee's facilities and deployment contexts annually.}''
\end{quote}

\subsection*{Practical Implementation Requirements}

Implementing veto clauses in licensing agreements requires:

\begin{itemize}
\item \textbf{Policy Development}: Universities must develop clear policies specifying which research triggers use restrictions and what those restrictions are.

\item \textbf{Technology Transfer Office Training}: TTOs must be trained to identify dual-use research and to incorporate veto clauses into licensing agreements.

\item \textbf{Legal Support}: Universities should establish legal capacity to draft and enforce use restriction clauses.

\item \textbf{Funding Agency Support}: Funding agencies can incentivize veto governance by conditioning grants on institutional establishment of use restriction policies.

\item \textbf{Researcher Communication}: Researchers should be informed that licenses will include use restrictions.
\end{itemize}

\subsection*{Limitations and Challenges}

University implementation faces several challenges:

\begin{itemize}
\item \textbf{Revenue Loss}: Broad use restrictions may reduce licensing revenue.

\item \textbf{International Complexity}: Technology transfer across national boundaries involves multiple legal jurisdictions.

\item \textbf{Monitoring Burden}: Verifying licensee compliance requires institutional capacity for monitoring and enforcement.

\item \textbf{Researcher Resistance}: Researchers may perceive use restrictions as constraints on research freedom.
\end{itemize}

\section*{C.2 Conference-Level Governance: Disclosure and Visibility}

\subsection*{Current Landscape}

Major AI conferences (NeurIPS, ICML, ICLR, ICRA, FAccT) serve as primary venues for research dissemination. Currently, conferences do not systematically assess dual-use risks or impose conditions on publication.

\subsection*{Mechanism: Dual-Use Disclosure and Use Conditions}

Conferences can implement lighter-touch governance by requiring disclosure of foreseeable dual-use risks and by enabling researchers to attach use conditions to published work:

\begin{enumerate}
\item \textbf{Dual-Use Disclosure Forms}: Conference submission requirements can include a ``dual-use and governance'' field where authors specify foreseeable military applications, circumstances under which they object to deployment, and recommended use conditions.

\item \textbf{Use Conditions and Licensing}: Research published at conferences can include explicit use conditions visible to readers. These conditions might specify: ``This research should not be deployed for autonomous weapons lacking meaningful human control.''

\item \textbf{Repository Integration}: Conference repositories can incorporate use condition metadata, making conditions machine-readable.

\item \textbf{Community Norms and Citation}: Conferences can establish professional norms that citing or building on research in violation of stated use conditions is ethically problematic.
\end{enumerate}

\subsection*{Implementation Requirements}

Conference-level implementation requires:

\begin{itemize}
\item \textbf{Policy Development}: Conference steering committees must develop policies specifying what dual-use disclosure requires.

\item \textbf{Review Procedures}: Reviewers and program committees need training to assess dual-use disclosure forms.

\item \textbf{Repository Infrastructure}: Conference repositories need technical capacity to incorporate use condition metadata.

\item \textbf{Community Communication}: Conferences should communicate to authors and readers that dual-use disclosure is expected.

\item \textbf{Transparency}: Policies and disclosure forms should be publicly available.
\end{itemize}

\section*{C.3 Funding Agency Requirements: Leverage Through Grant Conditions}

\subsection*{Current Landscape}

Funding agencies (NSF, DARPA, DOE, and international equivalents) distribute billions of dollars annually supporting research. Funding agencies exercise significant power over research directions through funding priorities and selection decisions. However, funding agencies rarely condition grants on governance requirements.

\subsection*{Mechanism: Governance Requirements as Grant Conditions}

Funding agencies can condition grants on institutional establishment and maintenance of veto governance structures:

\begin{enumerate}
\item \textbf{Dual-Use Research Identification}: Grant conditions can require that institutions identify research falling within designated dual-use categories and subject such research to veto governance.

\item \textbf{Veto Body Establishment}: Grant conditions can require that institutions establish veto committees with specified composition, documented procedures, and protected decision-making authority.

\item \textbf{Review Requirements}: Grants can require that dual-use research be subject to veto body review before publication, licensing, or partnership formation.

\item \textbf{Funding Ineligibility for Violations}: Grant conditions can specify that institutions violating veto governance lose eligibility for future funding.

\item \textbf{Transparency and Reporting}: Grants can require that institutions maintain public records of veto decisions.
\end{enumerate}

\subsection*{Implementation Example}

Suppose the National Science Foundation conditions grants on institutional governance:

\begin{quote}
``\textit{Institutions must establish a Dual-Use Research Governance Committee. Committee must include at least 2 faculty, 1 ethicist/policy expert, and 2 representatives from potentially affected communities. All research supported by this grant involving autonomous systems must be reviewed by the Committee before publication, licensing, or commercial partnership. Institutions must maintain public records of committee decisions.}''
\end{quote}

\subsection*{Advantages of Funding Agency Implementation}

Funding agency requirements offer several advantages:

\begin{itemize}
\item \textbf{Leverage}: Funding agencies exercise significant leverage over institutions through grant awards.

\item \textbf{Scale}: Conditions imposed by major funding agencies affect research across multiple institutions.

\item \textbf{Precedent}: Funding agencies have successfully imposed governance requirements in other domains (IRB requirements for human subjects research).

\item \textbf{Flexibility}: Funding agencies can develop requirements tailored to different research domains.
\end{itemize}

\subsection*{Limitations and Challenges}

Funding agency implementation faces challenges:

\begin{itemize}
\item \textbf{Political Opposition}: Requirements may face political opposition from researchers, universities, and defense contractors.

\item \textbf{International Coordination}: Without international coordination, national funding requirements become less effective.

\item \textbf{Non-Compliance and Evasion}: Institutions might nominally comply while maintaining governance structures incapable of actually constraining militarization.

\item \textbf{Institutional Autonomy}: Universities may resist funding conditions perceived as governmental overreach.
\end{itemize}

\section*{C.4 Repository-Level Artifact Governance: Infrastructure-Based Enforcement}

\subsection*{Current Landscape}

AI research artifacts---code, models, datasets---are increasingly released through online repositories: Hugging Face, GitHub, ModelHub, arXiv, Zenodo. These repositories serve as primary distribution channels for research artifacts. Repositories currently function as neutral distribution channels without systematic governance of downstream uses.

\subsection*{Mechanism: Machine-Readable Use Restrictions and Licensing}

Repositories can require that artifacts include machine-readable use restrictions specifying permitted and prohibited uses:

\begin{enumerate}
\item \textbf{License Standardization}: Artifacts can be released under licenses (based on GPL, MIT, or custom licenses) that explicitly specify use conditions. License text can be standardized using SPDX expressions.

\item \textbf{Metadata Embedding}: Models and datasets can include metadata fields specifying use conditions and governance requirements.

\item \textbf{Access Restrictions}: Repositories can implement access control tiers. Some artifacts might be available for research access, while deployment access requires additional authorization.

\item \textbf{Version Control and Update Requirements}: Repositories can require that organizations deploying artifacts under specified conditions maintain current versions.
\end{enumerate}

\subsection*{Mechanism: Audit and Delisting}

Repositories can implement compliance monitoring:

\begin{enumerate}
\item \textbf{Deployment Reporting}: Users accessing artifacts for deployment can be required to submit deployment context information.

\item \textbf{Audit Rights}: Repositories can assert rights to audit artifact uses, requiring deploying organizations to provide documentation.

\item \textbf{Violation Response}: If repositories determine that artifacts are being deployed in violation of use conditions, repositories can delist artifacts or revoke access.

\item \textbf{Public Transparency}: Repositories can maintain public records of use violations and enforcement actions.
\end{enumerate}

\subsection*{Advantages of Repository-Level Governance}

Repository governance offers advantages:

\begin{itemize}
\item \textbf{Infrastructure Leverage}: Repositories already control access and distribution.

\item \textbf{Technical Feasibility}: Machine-readable metadata and access control are established technical capabilities.

\item \textbf{Practical Friction}: Requiring users to certify compliance creates practical friction that increases consideration of restrictions.

\item \textbf{Scalability}: A single repository governance decision affects all users accessing that repository.

\item \textbf{Transparency}: Repository-level governance is relatively transparent.
\end{itemize}

\subsection*{Limitations and Challenges}

Repository-level governance faces challenges:

\begin{itemize}
\item \textbf{Enforcement Limits at National Borders}: Repositories cannot enforce restrictions on users outside their legal jurisdiction.

\item \textbf{Code Forking and Redistribution}: Repositories cannot prevent users from downloading artifacts, modifying them, and redistributing outside repository control.

\item \textbf{Scientific Knowledge Cannot Be Restricted}: Repositories can restrict specific code or model implementations but cannot restrict underlying scientific knowledge.

\item \textbf{Monitoring Burden}: Auditing deployments at scale requires significant resources.

\item \textbf{User Deception}: Organizations can misrepresent intended uses or use restricted artifacts while evading deployment reporting.
\end{itemize}

\section*{C.5 Community-Based Norms and Professional Solidarity}

\subsection*{Current Landscape}

Professional research communities have established ethics statements and codes of conduct. However, these statements typically lack enforcement mechanisms and are developed within elite academic institutions without meaningful participation from affected communities.

\subsection*{Mechanism: Professional Commitments and Collective Accountability}

Veto governance can be operationalized through professional norms and collective community commitments:

\begin{enumerate}
\item \textbf{Professional Codes of Conduct}: Research professional associations can establish binding codes of conduct specifying that members commit to refusing military funding or to imposing use restrictions on dual-use work.

\item \textbf{Institutional Commitments}: Research institutions can establish public commitments to governance of dual-use research, specifying that they will not accept certain types of military funding.

\item \textbf{Community Oversight and Accountability}: Research communities can establish transparency mechanisms where members can raise concerns about other members' research or institutional practices.

\item \textbf{Community-Led Governance Design}: Research communities should center affected communities in governance design, treating governance research as legitimate scholarly work.

\item \textbf{Sanctuary Provisions}: Research communities can establish commitments to protect researchers who refuse military funding.
\end{enumerate}

\subsection*{Advantages of Community-Based Governance}

Community-based approaches offer advantages:

\begin{itemize}
\item \textbf{Legitimacy from Within}: Governance emerging from research communities has legitimacy that external structures may lack.

\item \textbf{Flexibility and Adaptation}: Professional norms can be refined through community deliberation.

\item \textbf{Low Institutional Burden}: Does not require new governmental authority or institutions.

\item \textbf{Centering Affected Communities}: Professional norms can explicitly center affected communities.

\item \textbf{Peer Accountability}: Professional peers understand research and can make nuanced judgments.
\end{itemize}

\subsection*{Limitations and Challenges}

Community-based governance faces challenges:

\begin{itemize}
\item \textbf{No Enforcement Authority}: Absent institutional backing, professional norms are not binding.

\item \textbf{Freeriding and Competition}: Individual researchers or institutions that violate codes gain competitive advantages.

\item \textbf{Professional Heterogeneity}: Not all researchers identify with professional codes or associations.

\item \textbf{International Variance}: Professional norms vary across countries and research communities.

\item \textbf{Limited Reach to Practitioners}: Professional codes primarily constrain academic research.
\end{itemize}

\subsection*{Combining Implementation Pathways}

The five implementation pathways are most effective when combined. A comprehensive approach might involve:

\begin{itemize}
\item Funding agencies condition grants on institutional governance
\item Universities embed veto clauses in licenses and establish governance committees
\item Conferences require dual-use disclosure
\item Repositories implement use restriction metadata and compliance monitoring
\item Professional communities establish codes of conduct and protect researchers exercising veto authority
\end{itemize}


\section{DESIGN PRINCIPLES FOR LEGITIMATE VETO GOVERNANCE}

Veto power can be abused. Governance structures designed without safeguards against abuse can become tools for suppression of legitimate research, for politicization of governance, or for discrimination against vulnerable populations. This section articulates design principles that constrain veto authority while preserving its capacity to prevent militarization.

\section*{D.1 Scope Limitation and Predefined Thresholds}

\subsection*{Problem: Veto Without Boundaries}

Unconstrained veto authority is dangerous. If veto bodies can refuse research for any reason or can continuously expand the scope of research subject to veto, veto becomes a mechanism for suppression rather than constraint on militarization. Historical examples demonstrate this risk: governments have used ``national security'' justifications to suppress inconvenient research, to silence critics, and to prevent investigation of governmental wrongdoing.

\subsection*{Design Principle: Scope Limitation}

Legitimate veto governance must define veto authority narrowly. Veto should apply to specific downstream uses, not to research itself, publication, or inquiry:

\begin{enumerate}
\item \textbf{Downstream Use Limitation}: Veto authority should apply to transfer and deployment, not to research or publication. Researchers should retain freedom to conduct inquiries and to publish findings.

\item \textbf{Predefined Risk Categories}: Veto authority should apply only when research meets predefined risk criteria. Risk categories might include:
\begin{itemize}
\item Integration into autonomous weapons lacking meaningful human control
\item Deployment in mass surveillance targeting political opponents or ethnic minorities
\item Use in violation of international humanitarian law or human rights law
\item Deployment in systems designed to enable forced displacement or ethnic cleansing
\end{itemize}

\item \textbf{High Confidence Thresholds}: Veto should be triggered by credible, documented risks, not speculative or distant possibilities.

\item \textbf{Exemptions for Legitimate Uses}: Veto should explicitly exempt legitimate civilian uses even if technologies are theoretically dual-use.
\end{enumerate}

\subsection*{Implementation Example}

A veto body governing autonomous systems research might establish predefined thresholds:

\begin{quote}
``Veto Triggered If: Research has direct application to autonomous weapons lacking meaningful human control AND the research was funded by a defense agency OR the researcher has been approached by defense organizations about deployment OR the publication explicitly discusses weapons applications. Veto Not Triggered For: Research with general applicability to autonomous systems absent specific evidence of weapons application OR research with clear civilian applications absent evidence of military diversion.''
\end{quote}

\section*{D.2 Collective Decision-Making and Procedural Legitimacy}

\subsection*{Problem: Veto Concentrated in Individual Authority}

Veto power concentrated in individual authority (a single researcher, administrator, or official) is dangerous. Individuals may make decisions based on personal bias, political motivation, or institutional self-interest rather than on principled assessment of risk.

\subsection*{Design Principle: Collective Decision-Making}

Legitimate veto governance should make veto decisions through deliberative collective processes:

\begin{enumerate}
\item \textbf{Pluralistic Veto Bodies}: Veto decisions should be made by committees including researchers with technical expertise, ethicists, legal experts, and representatives from affected communities. Pluralistic composition ensures that decisions reflect multiple perspectives.

\item \textbf{Documented Procedures}: Veto bodies should operate according to documented procedures specifying how veto is initiated, what deliberation occurs, what information is reviewed, how decisions are made, and what records are maintained.

\item \textbf{Transparency and Publicity}: Veto decisions should be documented and made public (with appropriate protections for sensitive information).

\item \textbf{Deliberation Requirements}: Veto bodies should be required to articulate reasoning for veto decisions.
\end{enumerate}

\subsection*{Implementation Example}

A university veto committee charter might specify:

\begin{quote}
``\textit{Composition: 3 faculty researchers, 2 ethicists/policy experts, 2 community representatives, 1 legal expert. All members have equal voting authority. Decision Threshold: Veto requires 5 of 8 votes. Initiation: Veto can be initiated by any committee member, any faculty member, or external community representatives. Deliberation Timeline: 30 days from initiation. Documented Decision: Decisions recorded in writing including risk factors, harm prevention, and any conditions. Public Records: Decisions made public.}''
\end{quote}

\section*{D.3 Formal Appeal and Revisability}

\subsection*{Problem: Veto as Permanent Suppression}

Veto decisions that are permanent and unrevisable can become suppression mechanisms. If veto can never be overturned, veto transitions from governance to censorship.

\subsection*{Design Principle: Appeal and Revisability}

Legitimate veto governance should make veto decisions subject to appeal and revision:

\begin{enumerate}
\item \textbf{Appeal Procedures}: Veto decisions should be subject to appeal to a higher authority independent of the original veto body.

\item \textbf{Grounds for Appeal}: Researchers should be able to appeal on grounds including procedural error, factual disagreement, applicability of predefined thresholds, or proportionality.

\item \textbf{Conditional Veto}: Veto decisions can be conditioned on circumstances under which veto could be lifted.

\item \textbf{Sunset Provisions}: Veto decisions can include sunset provisions specifying that the veto will be revisited at a future date.

\item \textbf{Evidence-Based Revision}: If new evidence becomes available, veto decisions should be revised based on evidence.
\end{enumerate}

\subsection*{Implementation Example}

A university might establish an appeals procedure:

\begin{quote}
``\textit{First Appeal: Researchers can appeal veto committee decisions to a university appeals board including external experts and community representatives. The appeals board can reverse committee decisions if it finds procedural error or factual error. Sunset: All veto decisions are revisited annually. If circumstances have changed or evidence suggests veto is ineffective, the committee can revise decisions.}''
\end{quote}

\section*{D.4 Legal and Career Protections}

\subsection*{Problem: Retaliation Against Those Exercising Veto Authority}

If researchers and administrators exercising veto authority face retaliation, veto authority remains theoretical rather than practical.

\subsection*{Design Principle: Protected Refusal}

Legitimate veto governance must include legal and career protections:

\begin{enumerate}
\item \textbf{Anti-Retaliation Provisions}: Federal and institutional law should explicitly prohibit retaliation against researchers or administrators for exercising veto authority.

\item \textbf{Tenure Protections}: Tenured faculty exercising veto authority should be protected by tenure.

\item \textbf{Whistleblower Protections}: Researchers or administrators reporting violations of veto authority should be protected as whistleblowers.

\item \textbf{Indemnification}: Institutions and individuals should be indemnified against legal liability for good-faith exercise of veto authority.

\item \textbf{Funding Protection}: Funding agreements should specify that institutional compliance with veto governance does not affect grant funding.

\item \textbf{International Protection}: For researchers in countries with weak legal protection, international professional organizations and funding agencies should provide legal support.
\end{enumerate}


\section{ADDRESSING OBJECTIONS AND LIMITATIONS}

Veto governance will face objections from multiple directions. This section addresses common objections and acknowledges genuine limitations that veto governance does not resolve.

\section*{E.1 Political Capture and Weaponization}

\subsection*{Objection}

``Veto power will be captured by hostile actors and weaponized against legitimate research. Governments will use veto mechanisms to suppress research investigating governmental wrongdoing.''

\subsection*{Response}

This objection identifies a genuine risk. Governance mechanisms designed to constrain militarization could be misused to suppress inconvenient research. History provides examples: authoritarian governments have used ``national security'' justifications to suppress research on environmental contamination, governmental violence, and human rights abuses.

However, this risk must be weighed against the current absence of constraints on militarization. Currently, militarization proceeds largely unchecked. The choice is not between perfect governance and the status quo; the choice is between imperfect governance with some risk of abuse and no governance at all with certainty of ongoing militarization.

Several design features constrain risks of capture and abuse:

\begin{itemize}
\item \textbf{Narrow Scope}: Veto authority limited to specific military uses
\item \textbf{Predefined Triggers}: Veto triggered by specific, documented risks
\item \textbf{Pluralistic Decision-Making}: Veto decisions made by committees including multiple perspectives
\item \textbf{Independent Review}: Veto decisions subject to appeal
\item \textbf{Transparency}: Veto decisions documented and made public
\item \textbf{International Observers}: International observers in governance processes
\item \textbf{Sunset Provisions}: Veto decisions time-limited and revisited
\end{itemize}

\section*{E.2 Innovation Stalling and Research Slowing}

\subsection*{Objection}

``Veto power will slow research and reduce innovation. Researchers will hesitate to pursue dual-use work. The resulting slowdown will harm economic competitiveness and human welfare.''

\subsection*{Response}

This objection reflects a particular value ranking: that speed of research and technological capability are paramount. This is defensible but should be made explicit and should be subject to debate.

Several points address this objection:

\begin{itemize}
\item \textbf{Scope Limitation}: Veto applies only to downstream military use, not to research itself. Researchers remain free to conduct inquiries and publish findings.

\item \textbf{Modest Time Impact}: Even if veto review requires time, the time impact should be modest (30--60 days in research timelines spanning years).

\item \textbf{Selective Application}: Veto applies only to research with clear dual-use risks, not to mainstream AI research.

\item \textbf{Competitive Fairness}: If veto governance is applied universally, all researchers face equivalent constraints.

\item \textbf{Defensible Value Tradeoff}: There is defensibility to prioritizing safety and justice over speed. The post-World War II international community decided that preventing nuclear proliferation justified accepting slower development of nuclear energy technology.

\item \textbf{Harm Prevention Value}: Prevention of military AI integration creates real value: prevention of autonomous weapons reduces civilian casualties, prevention of surveillance systems targeting vulnerable populations protects autonomy and dignity.
\end{itemize}

\section*{E.3 Democratic Legitimacy of Veto Bodies}

\subsection*{Objection}

``Veto bodies will be undemocratic, wielding authority over research without legitimate democratic sanction. Who appointed these veto committees?''

\subsection*{Response}

This objection identifies a genuine tension: governance authority should be legitimate and accountable. Several design features address democratic legitimacy concerns:

\begin{enumerate}
\item \textbf{Community Leadership}: Veto bodies should be led and controlled by affected communities.

\item \textbf{Elected or Appointed with Accountability}: Veto body members can be elected by research communities or appointed through transparent processes with documented accountability.

\item \textbf{Appeal and Revision}: Veto decisions should be subject to appeal and revision.

\item \textbf{Public Participation}: Veto bodies should hold public meetings, accept public input, and maintain public records.

\item \textbf{Embedded in Democratic Institutions}: Veto governance should be embedded in democratic institutions (universities, funding agencies) that have governance mechanisms and accountability structures.

\item \textbf{Temporal Limits}: Veto body members should serve fixed terms with term limits.
\end{enumerate}


\section{MARGINALIZED COMMUNITY PERSPECTIVES}

Veto governance succeeds as governance only if it is designed with and controlled by those most harmed by military AI deployment. This section details why affected community leadership is essential and how governance should be structured to center community voice and authority.

\section*{F.1 Exclusion From Current AI Governance}

\subsection*{The Geography of AI Harm}

AI governance discourse is concentrated in Global North institutions. Yet the most severe harms of military AI deployment are borne in other regions. Autonomous weapons are tested and deployed in conflicts in the Middle East, East Africa, and Central Asia. Mass surveillance systems are deployed in countries with authoritarian governments. The populations experiencing these harms are largely excluded from AI governance design.

The exclusion of affected communities from governance design is both unjust and epistemically limiting. Affected communities possess detailed, situated knowledge about:

\begin{itemize}
\item How military AI systems are deployed and what harms result
\item What governance structures would be effective in their contexts
\item What unintended consequences governance mechanisms might have
\item How governance mechanisms intersect with other power dynamics affecting their communities
\end{itemize}

\subsection*{Epistemic Consequences of Exclusion}

Researchers and policymakers designing governance from elite institutions lack situated knowledge. Governance designed without affected community input risks being ineffective, culturally inappropriate, or inadvertently harmful.

\section*{F.2 Shifting Governance Leadership: Operationalizing Community Control}

\subsection*{Principle: Affected Communities Lead}

Justice in governance requires that affected communities control governance design and implementation. This is not merely about inclusion or consultation; it is about control.

\subsection*{Operationalization Requirements}

Shifting governance leadership to affected communities requires:

\begin{enumerate}
\item \textbf{Funding for Governance Research}: Governance research on military AI should be funded and led by researchers in affected regions. Funding structures should be reoriented to support governance research in affected regions.

\item \textbf{Treating Governance as Scholarship}: Governance research should be recognized as legitimate academic work. Researchers conducting governance work should be eligible for professorial positions and prestigious publication venues.

\item \textbf{Binding Partnerships}: Governance partnerships should establish binding decision authority for affected communities, not merely advisory roles.

\item \textbf{Resource Control}: Affected communities should control governance resources---funding, staffing, decision-making authority.

\item \textbf{International Structures}: International governance structures should be led by representatives from affected regions with binding authority.
\end{enumerate}

\section*{F.3 Militarization and Marginalization: Structural Interconnections}

\subsection*{Deep Linkages}

Militarization and marginalization are not separate problems; they are deeply interconnected. Military violence disproportionately targets marginalized populations. AI systems amplify these dynamics by automating and scaling targeting logics that already discriminate.

\subsection*{Targeting Dynamics}

Military violence targets populations perceived as threats. Populations perceived as threats are often marginalized groups: ethnic minorities, indigenous peoples, religious minorities, political dissidents. Military AI systems scale and automate these targeting dynamics.

Addressing militarization without centering marginalization treats symptoms rather than structural causes. Effective governance must address both military applications of AI and the way military AI automation amplifies existing targeting and marginalization.


\section{CASE STUDIES OF MILITARIZATION PATHWAYS}

This section details three historical cases where research transitioned into military applications, identifying veto points where governance could have constrained militarization.

\section*{G.1 Facial Recognition and Mass Surveillance}

\subsection*{Research Development}

Facial recognition technology developed over decades through open academic research. Major algorithmic innovations came from researchers in academia and industry, with algorithms published at conferences and in journals.

\subsection*{Military Pathways}

Facial recognition technology moved into military and surveillance contexts through multiple pathways:

\begin{enumerate}
\item \textbf{Surveillance Deployment}: Law enforcement and military organizations began deploying facial recognition systems.

\item \textbf{Border Control}: Border control and immigration enforcement began deploying facial recognition systems.

\item \textbf{Targeted Surveillance}: Authoritarian governments deployed facial recognition for targeted surveillance of ethnic minorities and political dissidents. Uyghur surveillance in Xinjiang relies heavily on facial recognition.

\item \textbf{Weapons Integration}: Facial recognition systems began to be integrated into weapons systems for target identification.
\end{enumerate}

\subsection*{Veto Points and Governance Failures}

Multiple veto points existed where governance could have constrained militarization:

\begin{enumerate}
\item \textbf{Publication Veto Point}: Individual papers could have been subject to veto review before publication or published with use restrictions.

\item \textbf{Licensing Veto Point}: When systems were licensed to commercial entities or governments, licensing agreements could have included restrictions on surveillance use.

\item \textbf{Partnership Veto Point}: University partnerships could have been subject to institutional review and could have been refused or conditioned.

\item \textbf{Funding Veto Point}: Government funding could have been conditional on governance requirements.
\end{enumerate}

\subsection*{Why Governance Failed}

Governance failed at each veto point for structural reasons:

\begin{itemize}
\item No publication review of dual-use risks
\item No technology transfer governance requiring use restriction assessment
\item No partnership review mechanism
\item No funding conditions on governance
\item Competitive pressure to publish rapidly, license broadly, and accept military funding
\end{itemize}

\section*{G.2 DARPA Funding in Autonomous Systems}

\subsection*{Research Development}

DARPA funds significant portions of AI research, particularly in autonomous systems, robotics, and sensor technology.

\subsection*{Military Pathways}

DARPA funding explicitly aims to develop military capabilities. Much DARPA research transitions directly into military systems:

\begin{enumerate}
\item \textbf{Autonomous Vehicles}: Research funded by DARPA has transitioned into autonomous military ground and aerial vehicles.

\item \textbf{Drone Control}: Research on drone control systems has transitioned into weapons systems.

\item \textbf{Robotic Manipulation}: Research on robotic systems has transitioned into military robotics.

\item \textbf{Autonomous Weapons}: Some DARPA funding explicitly aims to develop autonomous weapons capabilities.
\end{enumerate}

\subsection*{Veto Points and Governance Failures}

Veto points existed in DARPA funding decisions:

\begin{enumerate}
\item \textbf{Funding Decision Veto Point}: Universities could have refused DARPA funding or conditioned acceptance on governance requirements.

\item \textbf{Institutional Governance Veto Point}: Universities could have established governance committees reviewing DARPA-funded research before acceptance.

\item \textbf{Publication Veto Point}: Universities could have restricted publication of DARPA-funded research.

\item \textbf{Technology Transfer Veto Point}: Technology could have been restricted from transfer to weapons applications.
\end{enumerate}

\subsection*{Why Governance Failed}

Governance failed for structural reasons:

\begin{itemize}
\item Revenue dependence on research funding
\item No institutional governance structures reviewing defense funding
\item Normalization of military research funding
\item Competitive pressure to accept DARPA funding
\item Lack of institutional authority to refuse military funding
\end{itemize}

\section*{G.3 University--Defense Contractor Partnerships}

\subsection*{Partnership Development}

Universities frequently establish research partnerships with defense contractors (Lockheed Martin, Raytheon, Boeing, Northrop Grumman). These partnerships involve joint research, researcher visits, technology transfer agreements, and collaborative publication.

\subsection*{Military Pathways}

University-defense contractor partnerships transition research directly into weapons systems:

\begin{enumerate}
\item \textbf{Technology Transfer}: Technologies developed in partnerships are transferred to defense contractors and integrated into weapons systems.

\item \textbf{Researcher Expertise}: University researchers collaborate with defense contractors on weapons development.

\item \textbf{Credibility and Legitimacy}: Defense contractors benefit from association with university researchers and academic legitimacy.

\item \textbf{Talent Pipeline}: Partnerships create pipelines enabling university researchers to transition into defense contractor positions.
\end{enumerate}

\subsection*{Veto Points and Governance Failures}

Veto points existed for partnership decisions:

\begin{enumerate}
\item \textbf{Partnership Approval Veto Point}: Universities could have established governance committees reviewing proposed partnerships.

\item \textbf{Technology Transfer Veto Point}: Technologies could have been restricted from transfer to weapons applications.

\item \textbf{Researcher Participation Veto Point}: Researchers could have been prohibited from participating in weapons development partnerships.
\end{enumerate}

\subsection*{Why Governance Failed}

Governance failed for structural reasons:

\begin{itemize}
\item University partnerships provide funding and prestige
\item Researcher autonomy norms prioritize freedom to choose collaborators
\item No institutional governance structures reviewing partnerships
\item Defense contractor participation in university governance
\item Cold War legacy normalizing university-defense partnerships
\end{itemize}


\section{OPERATIONALIZATION MECHANISMS AND ENFORCEMENT STRUCTURES}

This section provides detailed mechanisms for operationalizing veto governance, addressing common concerns that veto governance lacks concrete implementation pathways.

\section*{H.1 Artifact-Level Triggers and Decision Protocols}

\subsection*{Problem: Vague Governance Without Decision Points}

Abstract veto authority is difficult to exercise. Without concrete decision points and specific procedures, veto authority remains theoretical.

\subsection*{Solution: Predefined Artifact-Level Triggers}

Veto governance becomes practical when embedded in concrete decision points where choices must be made.

\subsubsection{Grant Acceptance Trigger}

\textbf{Decision Point}: When institutions receive funding from defense agencies (DARPA, military research offices) or from other entities explicitly requesting dual-use research, a veto trigger activates.

\textbf{Required Process}:

\begin{enumerate}
\item \textbf{Written Risk Memo} (Responsibility: Technology Transfer Office or Grants Administrator)
\begin{itemize}
\item Identify foreseeable military integration pathways
\item Assess likelihood and severity of militarization
\item Document evidence supporting risk assessment
\item Specify conditions that might mitigate risks
\item Timeline: 10 days from grant notification
\end{itemize}

\item \textbf{Decision Deadline} (Responsibility: Veto Body)
\begin{itemize}
\item Veto body must convene and reach decision within 30 days
\item Failure to reach decision defaults to permission
\item Decision is recorded with reasoning
\end{itemize}
\end{enumerate}

\subsubsection{Licensing Terms Trigger}

\textbf{Decision Point}: When research is licensed to downstream organizations, a veto trigger activates.

\textbf{Required Process}:

\begin{enumerate}
\item \textbf{Written Risk Memo} (Responsibility: Technology Transfer Office)
\begin{itemize}
\item Identify proposed licensee and anticipated uses
\item Assess military applications likelihood
\item Specify what use conditions might prevent militarization
\item Timeline: 10 days from licensing proposal
\end{itemize}

\item \textbf{License Agreement Terms} (Responsibility: Technology Transfer Office with Veto Body Input)
\begin{itemize}
\item License must include use restrictions approved by veto body
\item License must require annual deployment reporting
\item License must grant audit rights
\item License must include indemnification clause for violations
\end{itemize}
\end{enumerate}

\subsubsection{Deployment Contracts Trigger}

\textbf{Decision Point}: When research transitions to deployment in actual use contexts, a veto trigger activates.

\textbf{Required Process}:

\begin{enumerate}
\item \textbf{Deployment Impact Assessment} (Responsibility: Deploying Organization and Licensor)
\begin{itemize}
\item Document deployment context
\item Assess alignment with use restrictions
\item Timeline: Assessment before deployment commences
\end{itemize}

\item \textbf{Veto Body Review}
\begin{itemize}
\item Assess whether deployment violates licensing terms
\item Approve, require modifications, or refuse deployment
\item Timeline: 30 days from assessment submission
\end{itemize}
\end{enumerate}

\subsubsection{Repository Releases Trigger}

\textbf{Decision Point}: When research artifacts are released to public repositories, a veto trigger activates.

\textbf{Required Process}:

\begin{enumerate}
\item \textbf{Use Condition Specification} (Responsibility: Authors/Repository Maintainers)
\begin{itemize}
\item Artifacts released with machine-readable metadata
\item License text specifies permitted and prohibited uses
\item Repository implements access controls enforcing restrictions
\end{itemize}

\item \textbf{Transparency and Monitoring} (Responsibility: Repository Platform)
\begin{itemize}
\item Repository maintains records of artifact access
\item Repository flags high-risk artifact deployments
\item Repository publishes aggregate compliance data
\end{itemize}
\end{enumerate}

\section*{H.2 Veto Body Composition: Structured Representation and Fairness}

\subsection*{Problem: Elite Capture and Marginalized Exclusion}

Without structured composition requirements, veto bodies risk becoming dominated by privileged actors while excluding affected communities and marginalized voices.

\subsection*{Solution: Fixed Representation Quotas and Support Structures}

\subsubsection{Representation Requirements}

A representative veto body of 8--10 members should have mandatory composition (see Table~\ref{table:veto_composition}):

\begin{table}[h]
\centering
\begin{tabular}{|l|c|l|}
\hline
\textbf{Member Category} & \textbf{Minimum} & \textbf{Selection Method} \\
\hline
Faculty Researchers (Technical) & 2--3 & Elected by research community \\
\hline
Ethicists / Policy Experts & 1--2 & Appointed by ethics/policy faculty \\
\hline
Affected Community Representatives & 2--3 & Appointed by affected communities \\
\hline
Legal Experts & 1 & Appointed by law faculty \\
\hline
External Experts (Optional) & 1--2 & Appointed by professional organizations \\
\hline
\end{tabular}
\caption{Mandatory veto body composition}
\label{table:veto_composition}
\end{table}

\textbf{Critical Principle}: All members have equal voting authority. Community representatives have full decision-making authority on par with faculty and experts.

\subsubsection{Community Representative Support Structures}

Material support for community participation:

\begin{enumerate}
\item \textbf{Stipends}: Community representatives receive substantial stipends (e.g., \$50--100/hour or annual stipends of \$5,000--10,000).

\item \textbf{Access to Independent Legal Counsel}: Community representatives can access independent legal counsel paid by the institution.

\item \textbf{Travel and Logistics Support}: All costs covered; institutions cover all logistics for community members.

\item \textbf{Training and Capacity Building}: Institutions provide training helping community representatives understand technical issues and governance procedures.

\item \textbf{Accessibility Accommodations}: Flexible meeting times, remote participation options, interpreters for language/disability access.
\end{enumerate}

\subsubsection{Term Limits and Rotation}

\begin{enumerate}
\item \textbf{Term Length}: Veto body members serve fixed terms (2--3 years).

\item \textbf{Term Limits}: Members can serve maximum 2 consecutive terms (4--6 years total).

\item \textbf{Staggered Rotation}: New members appointed on staggered schedules.

\item \textbf{Community Control of Appointments}: Affected community organizations control appointment of community representatives.
\end{enumerate}

\section*{H.3 Enforcement Mechanisms and Compliance Structures}

\subsection*{Problem: Veto Without Consequences}

Veto authority without consequences is unenforceable. Enforcement requires that violations trigger automatic consequences.

\subsection*{Solution: Multi-Layered Enforcement Architecture}

\subsubsection{Contractual Enforcement}

\textbf{Mechanism}: Research, licensing, and partnership agreements include explicit veto clauses creating legal obligations.

\textbf{Example Veto Clause Language}:

\begin{quote}
``Licensee agrees that:
\begin{itemize}
\item Research will not be deployed for autonomous weapons lacking meaningful human control; mass surveillance targeting political opponents; deployment in violation of international humanitarian law
\item Licensee will not sublicense to parties likely to use for prohibited purposes
\item Licensee will provide annual deployment context reports
\item Licensee will permit licensor to audit deployment contexts
\item Any violation entitles licensor to [remedies: license termination, indemnification for damages, injunctive relief]
\end{itemize}''
\end{quote}

\subsubsection{Machine-Readable Use Restrictions and Repository Enforcement}

\textbf{Mechanism}: Artifacts include machine-readable use restriction metadata that repositories can automatically enforce.

\textbf{Implementation}:

\begin{enumerate}
\item \textbf{License Metadata Standards}: Artifacts include metadata fields specifying use restrictions using standard formats (SPDX expressions, YAML metadata).

\item \textbf{Repository Enforcement}:
\begin{itemize}
\item Flag artifacts with use restrictions
\item Require users to certify compliance before downloading
\item Track which organizations access restricted artifacts
\item Block access by organizations with known violations
\item Delist artifacts if violations detected
\end{itemize}
\end{enumerate}

\subsubsection{Compliance Auditing and Reporting}

\textbf{Mechanism}: Organizations deploying research under use restrictions are subject to mandatory reporting and audit.

\textbf{Reporting Requirements}:

\begin{enumerate}
\item \textbf{Annual Deployment Reports}:
\begin{itemize}
\item Deployment location and organization
\item Intended use and actual use
\item Populations affected by deployment
\item Measures ensuring compliance
\item Modifications or updates to deployed systems
\end{itemize}

\item \textbf{Audit Notification}:
\begin{itemize}
\item Facility visits to assess deployment context
\item Document review (deployment logs, system configurations)
\item Staff interviews about deployment practices
\item Written audit report
\end{itemize}

\item \textbf{Public Reporting}: Licensor publishes aggregate data on compliance, deployment sectors, violation rates, and enforcement actions.
\end{enumerate}

\subsubsection{Funding Agency Enforcement}

\textbf{Mechanism}: Federal funding agencies make compliance with veto governance a condition of grant eligibility.

\textbf{Funding Conditions}:

\begin{enumerate}
\item \textbf{Institutional Governance Requirement}: Institutions must establish and maintain veto governance structures. Failure triggers loss of funding eligibility.

\item \textbf{Compliance with Veto Decisions}: Institutions must comply with veto decisions. Violations trigger investigation and potential defunding.

\item \textbf{Transparency Requirement}: Institutions must maintain public records of veto decisions.

\item \textbf{Loss of Funding for Non-Compliance}: Non-compliant institutions lose eligibility for future grants.
\end{enumerate}

\textbf{Example Federal Regulation}:

\begin{quote}
``\textit{Award recipients must establish institutional governance procedures for dual-use research. Research involving [specified dual-use categories] is subject to review by an institutional committee meeting criteria specified in this regulation. Institutions must comply with committee recommendations regarding use restrictions, licensing conditions, or research refusal. Institutions violating these requirements become ineligible for future awards until compliance is restored.}''
\end{quote}

\subsubsection{Indemnification and Legal Liability}

\textbf{Mechanism}: Violations of use restrictions trigger financial and legal consequences.

\textbf{Indemnification Clause Example}:

\begin{quote}
``If Licensee deploys research in violation of use restrictions and that deployment results in harm, Licensee indemnifies Licensor against:
\begin{itemize}
\item Damages paid to harmed parties
\item Legal costs and attorney fees
\item Institutional liability
\item Regulatory fines or sanctions
\item Reputational harm
\end{itemize}''
\end{quote}

\textbf{Legal Liability}:

\begin{enumerate}
\item \textbf{Civil Liability}: Deploying organizations can be sued for damages resulting from violations.

\item \textbf{Regulatory Liability}: Organizations face regulatory investigation and sanctions if deployment violates standards.

\item \textbf{Criminal Liability}: Criminal prosecution may be possible in some jurisdictions if violations result in harm.
\end{enumerate}

\section*{H.4 Integration with Existing Institutional Infrastructure}

\subsection*{Key Insight}

The machinery for veto governance already exists within institutional structures. Implementation requires embedding veto authority into existing decision-making processes.

\subsubsection{Universities: Repurposing Existing Authority}

\textbf{Technology Transfer Offices} already review licensing agreements and maintain licenses. \textbf{Mechanism}: TTO charters updated to require:
\begin{itemize}
\item Identification of dual-use research
\item Inclusion of use restrictions in licenses
\item Compliance monitoring and reporting
\item Authority to refuse licenses lacking adequate restrictions
\end{itemize}

\textbf{Faculty Governance Structures} already review research and establish institutional policies. \textbf{Mechanism}: Governance charters updated to:
\begin{itemize}
\item Require establishment of veto committee for dual-use research
\item Grant veto committee decision-making authority
\item Protect veto committee decisions from administrative override
\end{itemize}

\subsubsection{Funding Agencies: Leveraging Grant Conditions}

\textbf{Funding agencies} already establish grant conditions and monitor institutional compliance. \textbf{Mechanism}: Grant conditions updated to:
\begin{itemize}
\item Require institutional veto governance as condition of funding
\item Require compliance with veto governance decisions
\item Require transparency through public records
\end{itemize}

\subsubsection{Repositories: Leveraging Platform Infrastructure}

\textbf{Repositories} already manage metadata and control access. \textbf{Mechanism}: Repository platforms extend capabilities to:
\begin{itemize}
\item Include use restrictions in standardized metadata
\item Implement access controls based on use restrictions
\item Automate enforcement of machine-readable restrictions
\end{itemize}

\subsubsection{Professional Associations: Establishing Norms}

\textbf{Professional associations} already establish codes of conduct and run conferences. \textbf{Mechanism}: Associations:
\begin{itemize}
\item Establish explicit commitments to veto governance in codes of conduct
\item Require dual-use disclosure for conference papers
\item Require use condition specification for published research
\end{itemize}

\section*{Note}

Veto governance is not theoretical. The mechanisms detailed in this appendix show that veto governance can be operationalized through existing institutional structures, with concrete decision points, structured procedures, material enforcement mechanisms, and appropriate safeguards against abuse.

\textit{\textbf{Implementation requires political will, funding, and culture change.}} But the mechanisms are available. Universities can embed veto clauses in licenses. Funding agencies can condition grants on governance. Repositories can implement use restrictions. Professional communities can establish codes of conduct.

\textit{Affected communities must lead this transition.} Governance designed without affected community leadership risks failing to address actual harms and risks replicating power imbalances that produced militarization.

The operationalization mechanisms detailed in this appendix provide the practical pathway for that transition: from abstract commitment to concrete practice, from ethical exhortation to enforceable governance, from elite-designed governance to community-led accountability.

\section*{Author Contributions}
SS is the core contributor. AC gave inputs regarding Appendix.VJ and DC gave overall feedback.

\section*{Acknowledgments}
SS gratefully acknowledges Martian and Philip Quirke for their generous financial support of this work.

\end{document}

%% file: math_commands.tex
\usepackage{amsmath,amsfonts,bm}

\def\eqref#1{equation~\ref{#1}}

\def\1{\bm{1}}

\DeclareMathAlphabet{\mathsfit}{\encodingdefault}{\sfdefault}{m}{sl}
\SetMathAlphabet{\mathsfit}{bold}{\encodingdefault}{\sfdefault}{bx}{n}

